\documentstyle [11pt]{article}
\def\journal#1#2#3#4{\ {\sl #1}\ \underline{\bf #2}, {#3}\  ({#4})}

\def\AnnPhys{\journal{Ann.\ Phys.}}

\def\ibid{\journal{\em ibid.}}

\def\NPB{\journal{Nucl.\ Phys.\ {\bf B}}}

\def\Physica{\journal{Physica}}

\def\PhysRev{\journal{Phys.\ Rev.}}

\def\PRD{\journal{Phys.\ Rev.\ {\bf D}}}

\def\SovJNuclPhys{\journal{Sov.\ J.\ Nucl.\ Phys.}}

\def\ZPhys{\journal{Z.\ Phys.}}

\begin{document}

\newcommand{\be}{\begin{equation}}
\newcommand{\ee}{\end{equation}}
\newcommand{\barray}{\begin{eqnarray}}
\newcommand{\earray}{\end{eqnarray}}
\newcommand{\gn}{\mbox{$\gamma_{\stackrel{}{5}}$}}
\newcommand{\adag}{a^{\dagger}_{p,s}}
\newcommand{\atildedag}{\tilde{a}^{\dagger}_{-p,s}}
\newcommand{\bdag}{b^{\dagger}_{-p,s}}
\newcommand{\btildedag}{\tilde{b}^{\dagger}_{-p,s}}
\newcommand{\apsbeta}{a^{\beta}_{p,s}}
\newcommand{\apsbetadag}{a_{-p,s}^{\beta\dagger}}
\newcommand{\adagbdag}{a^{\dagger}_{p,s} b^{\dagger}_{-p,s}}
\newcommand{\aps}{a^{}_{p,s}}
\newcommand{\bps}{b^{}_{-p,s}}
\newcommand{\bpsbeta}{b^{\beta}_{p,s}}
\newcommand{\bpsbetadag}{b_{-p,s}^{\beta\dagger}}
\newcommand{\Adag}{A^{\dagger}_{p,s}}
\newcommand{\Bdag}{B^{\dagger}_{-p,s}}
\newcommand{\Aps}{A^{}_{p,s}}
\newcommand{\Apsbeta}{A^{\beta}_{p,s}}
\newcommand{\Apsbetadag}{A^{\beta\dagger}_{-p,s}}
\newcommand{\Bps}{B^{}_{p,s}}
\newcommand{\Bpsbeta}{B^{\beta}_{p,s}}
\newcommand{\Bpsbetadag}{B^{\beta\dagger}_{-p,s}}
\newcommand{\ApL}{A^{}_{p,L}}
\newcommand{\BpL}{B^{}_{-p,L}}
\newcommand{\ApR}{A^{}_{p,R}}
\newcommand{\BpR}{B^{}_{-p,R}}
\newcommand{\BdagL}{B^{\dagger}_{-p,L}}
\newcommand{\BdagR}{B^{\dagger}_{-p,R}}
\newcommand{\eps}{\epsilon}
\newcommand{\go}{\left( \gamma_o - \vec{\gamma} \cdot \hat{n} \right)}
\newcommand{\abab}{a^{\dagger}_{p,L}\,b^{\dagger}_{-p,L}\,a^{\dagger}_{p,R}
                   \,b^{\dagger}_{-p,R}}
\newcommand{\alphai}{\alpha_{i}}
\newcommand{\limit}{\lim_{\Lambda^2 \rightarrow \infty}}
\newcommand{\p}{\vec{p}, p_o}
\newcommand{\poprime}{p_o^{\prime}}
\newcommand{\prodps}{\prod_{p,s}}
\newcommand{\prodp}{\prod_{p}}
\newcommand{\psibar}{\bar{\psi}}
\newcommand{\psibarpsi}{ < \bar{\psi} \, \psi
            > }
\newcommand{\PsibarPsi}{ < \bar{\Psi} \, \Psi
            > }
\newcommand{\psibeta}{\psi^{}_{\beta}}
\newcommand{\psibarbeta}{\bar{\psi}_{\beta}}
\newcommand{\psidag}{\psi^{\dagger}}
\newcommand{\psidagbeta}{\psi^{\dagger}_{\beta}}
\newcommand{\psiL}{\psi_{_{L}}}
\newcommand{\psiR}{\psi_{_{R}}}
\newcommand{\Q}{Q_{_{5}}}
\newcommand{\Qa}{Q_{_{5}}^{a}}
\newcommand{\Qbeta}{Q_{5}^{\beta}}
\newcommand{\qqbar}{q\bar{q}}
\newcommand{\sumps}{\sum_{p,s}}
\newcommand{\TH}{\Theta}
\newcommand{\thetap}{\theta_{p}}
\newcommand{\costhetap}{\cos{\thetap}}
\newcommand{\sinthetap}{\sin{\thetap}}
\newcommand{\thetaset}{\{ \thetap  \}}
\newcommand{\thetapi}{\thetap{}_{i}}
\newcommand{\Tomega}{\frac{\Tprime}{\omega}}
\newcommand{\pomega}{\frac{p}{\omega}}
\newcommand{\Tprime}{T^{'}}
\newcommand{\Tprimesq}{T^{'2}}
\newcommand{\vac}{| vac \rangle}
\newcommand{\vacbeta}{| vac \rangle_{_{\beta}}}
\newcommand{\x}{\vec{x},t}
\newcommand{\xPrime}{\vec{x} - \hat{n} ( t - t'), t'}
\newcommand{\y}{\vec{y}, y_o}

\thispagestyle{empty}

\vspace*{-.15in}

\hspace*{\fill}\fbox{CCNY-HEP-94-5}

\begin{center}
{\Large {\bf Chiral Current at High Temperatures\fnsymbol{footnote}\footnote
{\samepage \sl
\noindent \parbox[t]{138mm}{ \noindent
                       Parts of this work have been supported in part by a
                       grant from NSF and from PSC-BHE of CUNY.
                           }

}
}
}\\
\baselineskip 5mm
\ \\
Ngee-Pong Chang (npccc@cunyvm.cuny.edu)\\
Department of Physics\\
City College \& The Graduate School of City University of New York\\
New York, N.Y. 10031\\
\  \\
Feb 2, 1994  \\
\end{center}

\vspace*{-.15in}

\noindent\hspace*{\fill}\parbox[t]{5.5in}{
        \hspace*{\fill}{\bf Abstract}\hspace*{\fill} \\
        {\em
        The BP-FTW effective action is a good laboratory
        to study the nature of chiral symmetry at high $T$.
        It is invariant under a new  Noether charge, $\Qbeta$.
        By explicit quantization, performed to order $\Tprime$
        we show that the thermal vacuum is annihilated
        by $\Qbeta$, so that the new chiral symmetry  is
        not broken spontaneously
        at high $T$.\\

        This high $T$ Noether charge is, however, different from the
        usual zero temperature chirality, $\Q$.  Our quantization
        also shows that the thermal vacuum is not annihilated by
        $\Q$, so that the old chiral symmetry remains broken
        at high $T$.\\

        The thermal vacuum is thus every bit
        as `structured' and `complex' as the
        Nambu-Jona-Lasinio vacuum
        at zero temperature.\\

        We exhibit the relation between the normal mode expansion
        for $\Qbeta$ versus the usual expansion for $\Q$. \\

        In an appendix, we point out the analyticity properties of the
        fermion propagator in real time formalism, and verify the
        canonical nature of the thermal Green function.
        }
                              }\hspace*{\fill} \\


\section{Introduction}

        It is well known that a massless quark propagates in the
        hot QCD (QED) environ as if it had a thermal mass
        $\propto gT$.$\cite{Weldon-Klimov}$
        But the hot QCD effective action$\cite{BP}$ for the quark propagation
        through the thermal medium is manifestly chiral
        invariant, so that the origin of this thermal mass remains
        a puzzle.
        At zero temperature, the origin of fermion mass is tied to the
        break down of chiral invariance.  At first sight, it would appear
        that at high $T$, the thermal mass has nothing to do with
        chirality.

        In QCD, it has long been known that $\psibarpsi$ vanishes
        above $T_c$.  The chiral symmetry of the hot QCD effective action
        would thus appear to be consistent with the usual conclusion on chiral
        symmetry restoration above $T_c$.

        And yet, the Noether current$\cite{Weldon-BP}$ associated with the
        new chiral symmetry at high $T$, $\Qbeta$,
        is different than the canonical $T=0$ chiral charge, $\Q$,
        and {\em  a priori} there is the possibility that the thermal
        vacuum at high $T$ continues to break $\Q$, even though
        it remains invariant under $\Qbeta$.

        In this paper, we perform a quantization of the BP-FTW
        effective action, and show how $\Qbeta$
        indeed leaves invariant the thermal vacuum at high $T$,
        while $\Q$ transforms it into a unitarily inequivalent new
        vacuum.

        The structure of the vacuum at high temperatures remains
        as complex as it was at zero temperature.  It belongs to
        a generalized class of NJL vacua.$\cite{NJL}$

\section{BP-FTW Action}

        In this paper, we shall be concerned with the BP-FTW action in the
        fermion sector.  It may be thought of as a
        generating functional for all the $n$-point functions involving
        two external fermion lines and any number of gluon lines. It
        may thus be used to describe gluon bremsstrahlung
        off an incident fermion in the hot environ.  To leading order in
        $T$, it is a gauge invariant action.

\be
   {\cal L}_{\rm eff} = - \bar{\psi}_{\beta} \gamma_{\mu} \partial^{\mu}
\psi_{\
                       + \frac{\Tprimesq}{2} \, \bar{\psi}_{\beta}
                         \left<
                       \frac{\gamma_o - \vec{\gamma} \cdot \hat{n} }
                       {D_o - \hat{n} \cdot \vec{D} }
                         \right> \psi_{\beta}          \label{BP-action}
\ee
     where
\begin{equation}
     \Tprimesq  = \frac{\textstyle g_r^2 C_f}{\textstyle 4} T^2
\end{equation}
     and $C_f$ is the Casimir invariant equal to $( N^2 - 1)/( 2 N )$
     for $SU(N)$ group.

        For our purposes here, we shall be interested in the properties
        of the vacuum in the absence of any external gluons, so that
        we may set the gluon field equal to zero and work with the
        action\footnote{
\samepage \sl
        In the real time formalism, there is a contribution  to the full
        action from the $\tilde{\psi}$ field associated with the heat
        bath.  It is given by the tilde operation$\cite{Umezawa}$ acting
        on the action in eq.(\ref{BP}), such that $I_{full} = I_{BP}
        \{ \psi \} - I_{BP} \{ \tilde{\psi} \}$
                       }
\be
        I_{\rm BP} = \int d^4 x \left\{
                  - \psibarbeta \gamma_{\mu} \partial^{\mu} \psibeta
                  - \frac{\Tprimesq}{4} \int dt' \left< \psibarbeta (\x)
                  \left( \gamma_o - \vec{\gamma} \cdot \hat{n} \right)
                  \psibeta(\xPrime) \right> \eps (t-t')
                             \right\}                 \label{BP}
\ee
        where the nonlocality of the effective action has been made
        manifest, and the angular bracket denotes an average over the
        orientation $\hat{n}$.

        The nonlocality of the BP-FTW action is notably different from
        the usual nonlocality we encounter in effective actions at zero
        temperature.  There
        the nonlocality is {\em  weak}, and may always be given in
        a derivative expansion.  Each term in the expansion is protected
        by the corresponding power of cutoff $\Lambda$ in the denominator.
        Such an expansion makes sense when we are talking about physics at
        a momentum scale below the cutoff.

        In our case here the nonlocality is proportional to $T^2$ in
        the numerator, and no derivative expansion is possible.

        The BP-FTW action is invariant under the global chiral transformation
\be
        \psibeta (\x)   \rightarrow  {\rm e}^{i \beta \gn}   \; \psibeta (\x)
\ee
        The Noether charge associated with this chiral symmetry may be
        obtained in the usual way resulting in
\barray
       \Qbeta &=&  \frac{1}{2} \;\; \int d^3 x \left\{ \; \psidagbeta
                        \, \gn \,
                        \psibeta - \frac{\Tprime {}^2}{8} \int dt_1 dt_2
                                   \; \epsilon (t_1 -t) \epsilon (t -t_2)
                                   \right.  \nonumber \\
             &&     \left. \, \psidagbeta (\vec{r} + \frac{\hat{n}}{2}
                                    (t_1 - t_2), t_1) \,
                    \left(  1 + \gamma_o \vec{\gamma} \cdot \hat{n} \right)\gn
\
                                     \psibeta (\vec{r} - \frac{\hat{n}}{2}
                                     (t_1 - t_2), t_2)
                                     \displaystyle
                                     \right\}    \label{Qbeta}
\earray
        In this paper, we wish to study the physics of this Noether charge.
        In particular, we wish to investigate if the new thermal vacuum
        is invariant under this Noether charge, and obtain the relation
        between $\Qbeta$ and the original $T=0$ Noether charge, $\Q$.

        To understand further the physics of this Noether charge, it is
        necessary to quantize the BP action.

        Because of the nonlocality, $\psibeta$ is not a canonical field
        in the BP action.
        That is, $\psidagbeta$ is not the canonical momentum for
        $\psibeta$. Nevertheless, as we shall show in the appendix, the
        vacuum  expectation of the equal time anticommutator obeys the
        usual rule
\be
               <  \{ \psibeta (\vec{x}, 0), \psidagbeta (\vec{y}, 0)  \}
>_{\be
               = \delta ( \vec{x} - \vec{y} )
\ee
        This comes about because by definition the thermal Green function
        is equal to the thermal average of the Heisenberg fields
\be
              < T \left( \psibeta (\x) \psibarbeta (0) \right)  >_{\beta}
              = \frac{1}{Z}\;  \sum_{n} {\rm e}^{-\beta E_n} \;
                         < n | T \left( \psi( \x) \psibar (0) \right)| n >
\ee
        so that at equal time, the vacuum expectation value of the
        anticommutator must reduce to the usual delta function.

\section{Quantization to order $\Tprime$}
        To quantize this action, it is convenient
        to work in momentum space, where the action takes the form
\be
        I_{\rm BP} = \int d^4 p \left\{
                      - i \psibarbeta (p) \left( \vec{\gamma} \cdot \vec{p}
                      - \gamma_o p_o \right) \psibeta (p)
                      + i \frac{\Tprime \stackrel{{}^2}{}}{2 \;\;}
                      \psibarbeta (p) \left( \vec{\gamma}\cdot \vec{p} \; a
                      - \gamma_o p_o \; b \right) \psibeta (p)
                                 \right\}
\ee
        with
\be
        a \equiv \frac{p_o}{2p^3} \left| \frac{p_o + p}{p_o - p}\right|
                 - \frac{1}{p^2}                     \label{fn-a-defined}
\ee
        and
\be
        b \equiv \frac{1}{2p p_o} \left| \frac{p_o + p}
                 {p_o - p}\right|                    \label{fn-b-defined}
\ee
        It is important to note that for all $p_o$ along the real axis,
        we are to take the principle values of the logarithms in the
        functions $a$ and $b$.  They satisfy the useful identity
\be
        p^2 a  \;-\; p_o^2 b   =  - 1                \label{ab-identity}
\ee

        We shall show in the appendix how this action will actually give
        rise to a fermion propagator $< T( \psibeta (\x) \psibarbeta (0))
        >_{\beta}$
        that has positive and negative energy poles along the real
        $p_o$ axis as well as a parallel pair of conjugate plasmino cuts
        in $p_o$ plane that extend from $-p$ to $p$ just above and below
        the real
        $p_o$ axis. The discontinuity across each cut is of order
        $\Tprime {}^2$.

        The poles include the pseudo-Lorentz invariant
        particle $\cite{Donoghue-Chang-hiT-Barton}$
        with  ( for $p$ in the range $ T >> p >> \Tprime$ )
\be
        p_o  =  \pm \left(  p + \frac{\Tprimesq}{2p}   + \ldots \right)
\ee
        with positive refractive index as well as positive residue.
        There are also the conjugate pair of poles
\be
        p_o  =  \pm ( p \pm \frac{2p}{e} {\rm e}^{- 4p^2 /\Tprimesq} )
\ee
        One of the two conjugate solutions has been dubbed the
        hole state by earlier analysis $\cite{Weldon-hole-1,Weldon-hole-2}$.
        It is a time like solution with $p_o^2 - p^2 > 0$.  This
        pole has positive residue, with however a {\em  negative} refractive
        index.  The other is a space like solution, with {\em  negative}
        residue.  Both solutions may be artifacts of the hard thermal
        loop perturbation theory.

        For our discussion here, we note that their residues are exponentially
        small (being of order ${\rm e}^{-4p^2 /\Tprimesq}$), so that they
        may be ignored in our quantization to order $\Tprime$.  In any
        case, as verified in the appendix, the two conjugate residues
        cancel out in the vacuum expectation value of the canonical
        anticommutation rule.

        Since we work only to order $\Tprime$, we may also ignore the
        $O(\Tprimesq/p^2)$ contributions due to the plasmino cut.
        This is valid for the range of $p$
\[                      \Tprime  \;<<\;    p   \;<<\; T
\]
        where $p$ is still `soft' relative to the hard thermal loop.

        The canonical field $\Psi$ may be obtained by a redefinition
        of $\psi$
\be
        \psi (p)  =  {\rm e}^{i \frac{\Tprime}{2} \TH} \; \Psi (p)
                     \sqrt{z_p}
\ee
        where $\TH \equiv \vec{\gamma}\cdot \vec{p} \;a \; - \gamma_{o} p_o
        \; b $, and $z_p$ is the wave function renormalization, which
        to order $\Tprime$ is simply unity.  With this
        field redefinition, we find
\be
        I_{\rm BP} = \int d^4 p \left\{
                      - \; i \; \overline{\Psi} (p) \left( \vec{\gamma}
                                \cdot \vec{p}
                      - \gamma_o p_o \right) \, \Psi (p)
                      - \Tprime \; \overline{\Psi}(p) \Psi (p)
                                 \right\}
\ee
        confirming that $\Psi$ indeed is the canonical massive Dirac field
        for the BP action.
        The $\Psi(\x)$ field has the canonical expansion$\cite{Chang-hisig}$
\be
        \Psi(\x) = \frac{1}{\sqrt{V}} \sumps {\rm e}^{i \vec{p} \,\cdot\,
                   \vec{x}}
                  \left\{  U_{ps} \,\Apsbeta {\rm e}^{- i\omega_p t}
                         + V_{ps} \,\Bpsbetadag {\rm e}^{+ i \omega_p t}
                  \right\}
\ee
        where the $\Apsbeta$ and $\Bpsbeta$ are
        the massive annihilation operators that satisfy the property
\be
        \Apsbeta \;| vac >_{_{\beta}} = \Bpsbeta   \;| vac >_{_{\beta}}
                        = 0
\ee
        To specify the $\pm i\eps$ boundary conditions,
        we require that at $t=0$, $\Psi$ field coincides with the free
        massless $\psibeta$ field in the interaction picture
\be
        \Psi ( \vec{x}, 0 ) = \psibeta ( \vec{x}, 0 )  \label{Psi-psi-relation}
\ee
        where
\be
        \psibeta(\vec{x},0) = \frac{1}{\sqrt{V}} \;\sumps {\rm e}^{i \vec{p}
                   \,\cdot\, \vec{x}}
                  \left\{  u_{ps} \,\apsbeta  \;+\; v_{ps} \,\bpsbetadag
                  \right\}                           \label{psi-beta-expansion}
\ee
        and $\apsbeta$ and $\bpsbeta$ are the original massless
        operators that annihilate the free thermal vacuum
\be
        \apsbeta | 0 >_{_{\beta}} = \bpsbeta |0> _{_{\beta}} = 0
\ee
        eq.(\ref{Psi-psi-relation}) may be solved in the usual way and
        leads to the familiar NJL relation
\barray
        \Apsbeta  &=&  \costhetap \apsbeta  \;+\; s \, \sinthetap
                       \bpsbetadag                   \label{Aps} \\
        \Bpsbeta  &=&  \costhetap \bpsbeta  \;-\; s \, \sinthetap
                       \apsbetadag                   \label{Bps}
\earray
        where
\be
        \tan{2\thetap}  = \frac{\Tprime}{p}
\ee
        and $s$ is defined to be $\pm 1$ for $R$ and $L$ helicities
respectively
        In real time formalism, the $\apsbeta$ and $\bpsbeta$ operators
        annihilate the free thermal vacuum, which itself includes the
        heat bath degrees of freedom.  If we represent the heat bath
        by tilde operators, we may relate the free thermal vacuum to the
        usual Fock space vacuum through the Bogoliubov transformation
        (as $T \rightarrow \infty$)
\be
        | 0 \rangle_{_{\beta}} = \prodps
        \left( \frac{1}{\sqrt{2}}-  \frac{1}{\sqrt{2}} \adag \atildedag \right)
        \left( \frac{1}{\sqrt{2}}+  \frac{1}{\sqrt{2}} \bdag \btildedag \right)
                                   | 0 \rangle
\ee
        and the new thermal vacuum of the hot QCD effective action
        is a generalized NJL vacuum
\be
        | vac \rangle_{_{\beta}} = \prodps
        \left( \cos{\thetap} - \;s \, \sin{\thetap} \;
                   a_{\beta,p,s}^{\dagger} b_{\beta,-p,s}^{\dagger} \right)
        \left( \cos{\thetap} - \;s \, \sin{\thetap} \;
               \tilde{a}^{\dagger}_{\beta,p,s}
               \tilde{b}^{\dagger}_{\beta,-p,s} \right)
                                   | 0 \rangle_{_{\beta}}
\ee
        where we have exhibited also the heat bath degrees of freedom
        in the thermal vacuum $| vac \rangle_{_{\beta}}$, and
        here $a_{\beta,p,s}^{\dagger},b_{\beta,p,s}^{\dagger} $ are
        the operators of the expansion in eq.(\ref{psi-beta-expansion}).

\section{New Chiral Charge}

        Having spelled out the relation between the massive Dirac field
        $\Psi$ and the perturbative $\psibeta$ in the interaction
        picture, we may finally express the two Noether charges directly
        in terms of their canonical annihilation and creation operators.
\be
        \Qbeta {}_{\rm full}   =  - \frac{1}{2} \;\sumps \; s \; \left(
                      \Adag \Aps + \Bdag \Bps
                      - \tilde{A}^{\dagger}_{p,s} \tilde{A}^{}_{p,s}
                      - \tilde{B}^{\dagger}_{p,s} \tilde{B}^{}_{p,s}
                                  \right)
\ee
        For comparison, we write down the canonical expansion for
        the $T=0$ Noether charge
\be
        \Q
        = - \frac{1}{2}
             \sumps \, s \; \left( \adag \aps +
             \bdag \bps - \tilde{a}_{p,s}^{\dagger} \tilde{a}_{p,s}^{}
             - \tilde{b}_{p,s}^{\dagger}
               \tilde{b}_{p,s}^{} \right)                   \label{Q5}
\ee
        and we see that the new Noether charge $\Qbeta$ is a
        very subtle replacement of the old massless operators
        with the new massive ones.

        It is apparent that
\be
        \Qbeta   | vac \rangle_{_{\beta}}  = 0
\ee
        while
\be
        \Q    | vac \rangle_{_{\beta}} \neq  0
\ee
        These results clearly show that the new chiral symmetry
        is not spontaneously broken at high $T$.  But the thermal
        vacuum is every bit as `structured' and `complex' as the
        Nambu-Jona-Lasinio vacuum at zero temperature.  The
        old chiral charge $\Q$ does not annihilate the thermal
        vacuum, but rotates it into another unitarily inequivalent
        generalized NJL vacuum. Dynamical symmetry breaking
        effects continue to dress the Fock space vacuum into the full
        fledged quark-antiquark paired ground state even
        at high $T$, and if anything the heat bath is a good
        source of additional quarks or antiquarks to form the
        pairing.

        In this paper, we have carried out the quantization of the
        BP-FTW action to order $\Tprime$ and found an interesting
        dichotomy between the new chiral charge at high $T$ and
        the old zero temperature chirality.
        Implications of this for the state of chiral symmetry in
        hot QCD will be reported elsewhere.$\cite{Chang-Banff}$

\section{Acknowledgments}

        This work was done while the author was a
        visitor at the Institue of Physics, Academia Sinica, Taipei,
        Republic of China.  The author wishes to thank Prof Shih-Chang
        Lee and the theory group for the warm hospitality, and to
        especially thank Prof H.L. Yu and Dr. B. Rosenstein for conversations.

\section*{Appendix: Analyticity Properties}

        In this appendix, we display the analyticity properties of the
        fermion propagator, $S_{\beta}(\p)$,$\cite{Weldon-hole-1}$
\be
        \langle  T(\psi(\x) \, \psibar(\y)) \rangle_{\beta}
                 \equiv  \int \frac{d^4 p}{ (2\pi)^4} {\rm e}^{ i
                 p \cdot (x - y) }
                 S_{\beta} ( \p )
\ee

        In the presence of interactions, the thermal propagator has the
        decomposition
\be
        S_{\beta} ( \p )
                 =  +  \vec{\gamma} \cdot \vec{p} \;s_{_{3}}(p, p_o, T)
                    -  \gamma_o p_o \;s_{o} (p, p_o, T)
                                                        \label{prop-2}
\ee
         where

\be
        s_{j} (p, p_o, T) = \int_{-\infty}^{\infty} \frac{dp_o^{\prime}
                    \rho_{j} (p, p_o^{\prime}, T)}
                    {p_o - p_o^{\prime} + i\eps}
                    + \int_{-\infty}^{\infty} \frac{dp_o^{\prime}
                    {\rm e}^{-\beta p_o^{\prime}} \rho_{j} (p, p_o^{\prime},
T)}
                    {p_o - p_o^{\prime} - i\eps}
                                                      \label{sj-rep}
\ee

        The $j=0$ spectral function is positive in the sense that
\be
         \omega \; \rho_o (p, \omega, T) \geq 0  \;\;\;\;{\rm for \;\; all
                                               \;\; real \;\;}\omega
\ee
        and satisfies the sum rule
\be
        \int_{-\infty}^{\infty} d\omega \; 2 \omega \, \rho_{o} ( p, \omega,
             T)  = \int_{o}^{\infty} d\omega \; 2\omega
             \overline{\rho}_{o} ( p, \omega, T) = 1     \label{rho0-sum-rule}
\ee
        for all $p$ and $T$.
        In this appendix we shall directly verify the sum rule in our
        perturbative calculation to order $\Tprimesq$
        for all $p$ in the range $ T >> p >> \Tprime$.
        The sum rule ensures the canonical
        anticommutation rule for the Heisenberg fields and enables us
        to carry out the quantization discussed in this paper.

        $\rho_{_{3}}$ is not positive definite, but nevertheless satisfies
        the inequality
\be
        (p_o \;\rho_{_{o}})^2 \geq (p \;\rho_{_{3}})^2
\label{rho3-inequality}
\ee
        Here $\overline{\rho}_{j}$ is related to $\rho_{j}$ by
\be
        \rho_{j} ( p, p_o^{\prime}, T ) =
             \frac{1}{ 1 + {\rm e}^{- \beta p_o^{\prime}}}
             \overline{\rho}_{j} ( p,
             p_o^{\prime}, T )                       \label{rhobar-def}
\ee

        We note in passing that in view of the positivity of the spectral
        function, eq.(\ref{sj-rep}) shows that $s_o$ function is
        actually a Herglotz function in the complex
        $p_o^2$-plane.$\cite{Herglotz}$
        As a consequence, the poles of the Herglotz function along the
        real axis must all have positive residues.

        From the representation eq.(\ref{sj-rep}) we also see that
        in the limit of high $T$, {\em the propagator functions $s_j$
        are real all along the real $p_o$-axis}.  This reinforces the
        earlier remark after eq.(\ref{fn-a-defined},\ref{fn-b-defined})
        that we are to take the principal values of the logarithms involved
        for $p_o$ along the entire real axis.

        With this in mind, we may go back to the BP-FTW action and
        write down by inspection the inverse propagator functions
\be
        S^{-1}_{\beta} (\p)  \equiv \left( - \vec{\gamma} \cdot \vec{p} r
                       + \gamma_o \, p_o \, r \kappa \right)
\ee
        where
\barray
        r  &=& 1 - \frac{\Tprimesq}{2} a  \\
        r \kappa &=& 1 - \frac{\Tprimesq}{2}  b
\earray
        and $\kappa$ has the physical interpretation of the index of
        refraction for the quark propagation through the hot medium.

        The propagator functions $s_j$, for $p_o$ real, may be
        simply expressed
\barray
        s_o ( p, p_o, T) &=& - \frac{1 - \frac{\Tprimesq}{2} b }{p^2 +
\Tprimesq
                           - p_o^2  + \frac{\Tprime {}^4}{4} (p^2 a^2
                           - p_o^2 b^2)  }           \label{so-rep-1}\\
        s_{_{3}} (p, p_o, T) &=& - \frac{1 - \frac{\Tprimesq}{2} a }{p^2
                           + \Tprimesq  - p_o^2
                           + \frac{\Tprime {}^4}{4} (p^2 a^2
                           - p_o^2 b^2)  }           \label{s3-rep-1}
\earray
        Written in this way, for $p >> \Tprime$ it becomes manifest that
        there is a pole at
\be
        p_o = \pm \omega_p \equiv \pm \left( p  +  \frac{\Tprimesq}{2p}  -
\frac
                            \ln{ \frac{4p^2}{\Tprimesq} } + \ldots  \right)
\ee
        with residue, $Z_p$, so that in the neighborhood of the pole we
        have
\be
        s_o ( p, p_o, T ) =  \frac{Z_p}{(2\omega_p)(p_o - \omega_p)} + \ldots
\ee
        and
\be
        Z_p =   \frac{\omega_p^2 - p^2}{\Tprimesq} = 1 - \frac{\Tprimesq}{4p^2}
                (\ln{ \frac{4p^2}{\Tprimesq} } - 1 ) + \ldots
\ee
        This pole is of course the expected radiative correction to the free
        particle pole $p_o = p$.
        Eq.(\ref{so-rep-1}), however, also shows a pair of poles at
        ( $p >> \Tprime$ ) for both positive and negative $p_o$
\be
        p_o = \pm \omega_{h\pm}  \equiv  \pm \left( p  \pm \frac{2p}{e}
                           {\rm e}^{-4 p^2/\Tprimesq} \right) + \ldots
\ee
        with residues
\be
        Z_{h\pm}   =  \pm  \frac{4 p^2}{\Tprimesq \, e}
                      {\rm e}^{- 4p^2/\Tprimesq}     \label{Z-hole}
\ee
        The solution $\omega_{h+}$ has been studied extensively by
        Weldon$\cite{Weldon-hole-1,Weldon-hole-2}$ but $\omega_{h-}$
        has apparently not been noticed.  It is a solution to
        the equation
\be
        p^2 r^2 - p_o^2 r^2 \kappa^2  = 0
\ee
        for $p_o$ in the space like region.  As mentioned earlier, the
        $r$ and $r\kappa$ functions are defined for all values of
        $p_o$ by taking the principal part of the logarithms, as
        instructed by the analyticity representation of eq.(\ref{sj-rep}).

        The residue for the new $\omega_{h-}$ hole solution is negative
        thus violating the positivity requirement of general field theory.
        The conjugate $\omega_{h+}$ hole solution has {\em  negative} index
        of refraction ($\kappa < 0$).  Since at $T=0$, the index of
        refraction was unity, this means that at some intermediate $T$
        the wave velocity of the quark in the medium became infinity.
        Both features could be signs that the so-called hole states are an
        artifact of perturbation theory.
        They arise due to the absence of higher order radiative corrections
        in the BP-FTW action.  Some have argued that there are no
        higher order radiative corrections to the BP-FTW action due
        to hard thermal loops.  If so, then we are faced with a dilemma
        between perturbation theory and general positivity requirements
        of field theory.

        Fortunately for us, if we work to order $\Tprimesq$, we do not
        have to face up to this problem.  The extra hole-state solutions
        are associated with the $\Tprime {}^4$ term in the denominator.
        We have verified explicitly to order $\Tprimesq$ that the propagator
        nevertheless satisfies the sum rule, assuring us of the canonical
        anticommutation rule at equal time.

        In continuing eq.(\ref{so-rep-1},\ref{s3-rep-1}) into the complex
        $p_o$-plane, we also meet with cuts due to the logarithms.
        To keep the reality condition, there are a mirror pair of cuts
        parallel to the real axis,
        both extending between $p_o = -p$ and $p_o = p$ just above and
        below the real axis.   In this way, the logarithms are real
        over the entire $p_o$ real axis.
        With this, we are finally ready to write down the complete
        analytic representation for $s_o$
\barray
        p_o \,s_o (p, p_o, T) &=&  \frac{1}{2} \sum_i \frac{Z_i}{p_o - \omega_i
                                 + \frac{1}{2} \sum_i \frac{Z_i}{p_o + \omega_i
                        && + \int_{-p}^{+p} d\poprime \frac{\poprime \rho_o}
                             {p_o - \poprime + i \eps} +
                             \int_{-p}^{+p} d\poprime \frac{\poprime
                             {\rm e}^{-\beta \poprime} \rho_o}
                             {p_o - \poprime - i \eps}
\earray
        where
\be
        \poprime \rho_o (p, \poprime, T) = \theta ( p^2 - \poprime {}^2 )
                 \left( \frac{\Tprimesq}{8 p}
                 \frac{1}{p^2 + \Tprimesq - \poprime {}^2}
                - \frac{\Tprime {}^4}{8 p^3} \frac{1}{(p^2 + \Tprimesq
                  - \poprime {}^2)^2}  \right)
\ee
        and we have only kept terms that contribute in the
        leading $\Tprimesq / p^2$ order.

        In this form, it is easy to verify analytically, to leading
        order in $\Tprimesq /p^2$ that eq.(\ref{rho0-sum-rule})
        is indeed satisfied.  As mentioned earlier, on account of
        eq.(\ref{Z-hole}), the pair of hole
        states $\omega_{h\pm}$ cancel out in the sum rule.


\begin{thebibliography}{99}
\bibitem{Weldon-Klimov}
              H.A. Weldon, \PhysRev {D26} {2789} {1982}; V.V. Klimov,
              \SovJNuclPhys {33}{934} {1981}.
\bibitem{BP}
              J.C.Taylor and S.M.H.Wong, \NPB{346}{115}{1990};
              E. Braaten and R. Pisarski, \PhysRev {D45}{1827}{1992};
              J. Frenkel and J.C. Taylor, \NPB{374}{156}{1992}.
\bibitem{Weldon-BP} A.H. Weldon, Proceedings of Winnipeg Summer School 1992,
              Canadian J. Phys.  See also J.P. Blaizot, E. Iancu, {\em Soft
              Collective Excitations in Hot Gauge Theories},
SACLAY-SPHT-93-064,
              Jun 93.
\bibitem{NJL}
              Y. Nambu and G. Jona-Lasinio, \PhysRev {122}
              {345} {1961}; \ibid {124} {246} {1961}.
\bibitem{Umezawa}
              H. Umezawa, H. Matsumoto, M. Tachiki, {\em  Thermo Field
              Dynamics and Condensed States}, N. Holland, Amsterdam,(1982).
\bibitem{Donoghue-Chang-hiT-Barton}
              J.F. Donoghue and B.R. Holstein, \PhysRev  {D28}
              {340} {1983}; \ibid {29}{3004 (E)} {1984};
              L.N. Chang, N.P. Chang, K.C. Chou,
              \PhysRev {D43}{596} {1991}.  See also
              G. Barton, \AnnPhys {200}{271} {1990}.
              In contrast with ref. \cite{Weldon-Klimov}, the authors
              here take the perturbative approach and regard $\Tprime$ as
              a small parameter.
              The difference is negligible for $T >> p >> \Tprime$
              in the range where $p$ is still `soft'.
\bibitem{Weldon-hole-1}
       H.A. Weldon, \PRD {40} {2410} {1989}; \Physica {A158} {169} {1989}.


\bibitem{Weldon-hole-2}
       G. Baym, J.P. Blaizot, B. Svetitsky, \PRD {46}{4043}{1992}.
       E. Petitgirard, \ZPhys {C54}{673} {1992}.
\bibitem{Chang-hisig} L.N. Chang, N.P. Chang, K.C. Chou,
	      {\em Phys. Rev.} \underline{D43}, 596 (1991).
\bibitem{Chang-Banff}
              N.P. Chang, ``Braaten-Pisarski Action, Disoriented
     Chiral Condensate, and Chiral Symmetry Non-Restoration'',
     Proceedings of 3rd Thermal Fields Workshop, Aug 16-27, 1993,
     Banff, Canada, World Sci. Publishers (Singapore), 1994.

\bibitem{Herglotz}
       See K. Nishijima, {\em  Fields and Particles}, W.A. Benjamin, New
       York,(1969), p. 449. \\
       See also S. Weinberg, \PhysRev {124} {2049} {1961}.





\end{thebibliography}
\end{document}